\pdfoutput=1

\documentclass[prl,twocolumn,showpacs,preprintnumbers,amsmath,amssymb,
superscriptaddress,nofootinbib]{revtex4-1}
\usepackage{epsfig}
\usepackage{graphicx}

\newcommand{\smallminus}{{\rm\rule[2.4pt]{6pt}{0.65pt}}}
\newcommand{\smallplus}{\hspace{0.5pt}\text{{\small+}}\hspace{-0.5pt}}
\newcommand{\mi}{\smallminus}
\newcommand{\psl}{\smallplus}

\begin{document}

\preprint{PUPT-2436}

\title{Recursion Relations for Tree-level Amplitudes in the $SU(N)$ Non-linear Sigma Model}

\author{Karol Kampf}
\affiliation{Institute of Particle and Nuclear Physics, Charles
University in Prague, Czech Republic}

\author{Jiri Novotny}
\affiliation{Institute of Particle and Nuclear Physics, Charles
University in Prague, Czech Republic}

\author{Jaroslav Trnka}
\affiliation{Institute of Particle and Nuclear Physics, Charles
University in Prague, Czech Republic}
\affiliation{Department of
Physics, Princeton University, Princeton, NJ, USA}

\date{\today}

\begin{abstract}
It is well-known that the standard BCFW construction cannot be used
for on-shell amplitudes in effective field theories due to bad
behavior for large shifts. We show how to solve this problem in the
case of the $SU(N)$ non-linear sigma model, i.e. non-renormalizable
model with infinite number of interaction vertices, using scaling
properties of the semi-on-shell currents, and we present new
on-shell recursion relations for all on-shell tree-level amplitudes
in this theory.
\end{abstract}

\maketitle

\section{Introduction}

Scattering amplitudes are physical observables that describe
scattering processes of elementary particles. The standard
perturbative expansion is based on the method of Feynman diagrams.
In last two decades there has been a huge progress on alternative
approaches, driven by the idea that the amplitude should be fully
determined by the on-shell data with no need to access the off-shell
physics. This effort has lead to amazing discoveries that have
uncover many surprising properties and dualities of amplitudes in
gauge theories and gravity. One of the most important breakthroughs
in this field was the discovery of the BCFW recursion relations
\cite{Britto:2004ap,Britto:2005fq} that allow us to reconstruct the
on-shell amplitudes recursively from most primitive amplitudes. They
are applicable in many field theories, however, in some cases like
effective field theories they can not be used.

Effective field theories play important role in theoretical physics.
One particularly important example is the $SU(N)$ non-linear sigma
model which describes the low-energy dynamics of the massless Goldstone bosons
corresponding to the chiral symmetry
breaking $SU(N)\times SU(N)\rightarrow SU(N)$. In the low energy QCD
they are associated with the octet of pseudoscalar mesons and the
model provides leading order predictions of
interactions of pions and kaons that dominate hadronic world at
lowest energies. It is also a starting point for many extensions or
alternatives of electroweak standard model.

In this short note we find the recursion relations for all
tree-level amplitudes of Goldstone bosons in the $SU(N)$ non-linear
sigma model. The importance of this result is two-fold: (i) It shows
that the BCFW-like recursion relations can be applicable to much
larger class of theories than expected before. This might also help
to understand better properties of the theory invisible otherwise.
It also tells us that the $SU(N)$ non-linear sigma model despite
being an effective (and therefore non-renormalizable) field theory
behaves in some cases similar to renormalizable theories. (ii) It
provides an effective tool for leading order (tree-level)
calculations of amplitudes with many external pions which might be
important for low energy particle phenomenology. More detailed
description together with other results will be presented in
\cite{future}.

\section{BCFW recursion relations}

Let us consider an $n$-pt on-shell scattering amplitude of massless
particles, and denote $t^a$ the generators of the Lie algebra of
corresponding global symmetry group $G$. If at tree-level each Feynman diagram
carries a single trace ${\rm  Tr}(t^{a_1}t^{a_2}\dots
t^{a_n})$, we can decompose the full amplitude  ${\cal A}_ n$ into
sectors with the same group factor,
\begin{equation}
{\cal A}^{tree}_n = \sum_{\sigma/{\bf Z}_n} A_n(p_{\sigma(1)},\dots
p_{\sigma(n)}) {\rm Tr}(t^{\sigma(1)}\dots
t^{\sigma(n)})\,,\label{color}
\end{equation}
where the sum is over all non-cyclic permutations. For each {\it
stripped} amplitude $A_n$ we have a natural ordering of momenta
$p_{\sigma(1)},\dots p_{\sigma(n)}$ and a single term
$A_n(p_1,p_2,\dots p_n)$ generates all the other by trivial
relabeling. At the loop level we can define analogous object in the
planar limit but in the general case this simple decomposition is
not possible due to terms with multiple traces.

In 2004 Britto, Cachazo, Feng and Witten (BCFW)
\cite{Britto:2004ap,Britto:2005fq} found a recursive construction of
tree-level on-shell amplitudes. The stripped amplitude
$A_n=A_n(p_1,\dots p_n)$ is a gauge invariant object and one can try
to fully reconstruct it from its poles. Because of the ordering the
only poles that can appear are of the form $P_{ab}^2 =0 $ where
$P_{ab} = \sum_{k=a}^bp_k$ for some $a$,$b$. On the pole the
amplitude factorizes into two pieces,
\begin{equation}
A_L (p_a,\dots p_b,-P_{ab})
\frac{1}{P_{ab}^2} A_R (P_{ab},p_{b\psl1},\dots
p_{a\mi1})\,.\label{fact}
\end{equation}
Let us perform the following shift on the external data:
\begin{equation}
p_i(z) = p_i + zq, \qquad p_j(z) = p_j - zq\,,
\end{equation}
where $i$ and $j$ are two randomly chosen indices, $z$ is a complex
parameter and $q$ is a fixed null vector which is also orthogonal to
$p_i$ and $p_j$, $q^2=(q\cdot p_i)=(q\cdot p_j)=0$. Note that the
shifted momenta remain on-shell and still satisfy momentum conservation. The
original
amplitude $A_n$ becomes a meromorphic function $A_n(z)$ with only simple
poles and if it
vanishes for $z\rightarrow\infty$ we can use Cauchy theorem to
reconstruct it,
\begin{equation}
A_n(z) = \sum_i \frac{{\rm Res}(A_n,z_i)}{z-z_i}\,,\label{Cauchy}
\end{equation}
where $z_i$ are poles of $A_n(z)$,
\begin{equation}
P_{ab}(z)^2 = (p_a+\dots + p_i(z)+\dots p_b)^2 = 0\label{Zpole}\,,
\end{equation}
located in $z_{ab}=-P_{ab}^2/2(q\cdot P_{ab})$. Note that $A_n(z)$
has a pole only if $i\in(a,\dots b)$ or $j\in(a,\dots b)$ (not both
or none). There exists a convenient choice $j=i\psl1$ which
minimizes a number of terms in (\ref{Cauchy}). According to
(\ref{fact}) ${\rm Res}(A_n,z_i)$ is a product of two lower point
amplitudes with shifted momenta and the Cauchy theorem
(\ref{Cauchy}) can be rewritten as
\begin{equation}
A_n(z) = \sum_{a,b} A_L(z)\frac{1}{P_{ab}^2}A_R(z)\,,\label{bcfw}
\end{equation}
where the sum is over all poles $P_{ab}(z)^2=0$ and
\begin{align}
A_L(z) &= A_L(p_a,\dots,p_i(z),\dots p_b,P_{ab}),\\
A_R(z) &= A_R(-P_{ab},p_{b\psl1},\dots,p_j(z),\dots p_{a\mi1}).
\end{align}

In the physical case we set $z=0$. $A_L$ and $A_R$ in (\ref{bcfw})
are lower point amplitudes, $n_R,n_L<n$ and therefore we can
reconstruct $A_n(z)$ recursively from simple on-shell amplitudes not
using the off-shell physics at any step. BCFW recursion relations
were originally found for Yang-Mills theory
\cite{Britto:2004ap,Britto:2005fq}, and proven to work in gravity
\cite{Bedford:2005yy,Cachazo:2005ca}. There are many works showing
validity in other theories (e.g. for coupling to matter see
\cite{Cheung:2008dn}).

If the amplitude $A_n(z)$ is constant or grows for large $z$, the
prescription (\ref{Cauchy}) cannot be used directly. The constant
behavior was studied e.g. in \cite{Feng:2009ei} on the cases of
$\lambda\phi^4$ and Yukawa theory. In the generic situation of a
power behavior $A_n(k)\approx z^k$ for $z\rightarrow\infty$ we can
use the following formula \cite{future}
\begin{align}
A_n(z)=\sum_{i=1}^{n}\frac{\mathrm{Res}\left( A_n;z_{i}\right) }{z-z_{i}}%
\prod_{j=1}^{k+1}\frac{z-a_{j}}{z_{i}-a_{j}}
\notag\\
+\sum_{j=1}^{k+1}A_n(a_{j})%
\prod_{l=1,l\neq j}^{k+1}\frac{z-a_{l}}{a_{j}-a_{l}}\,,
\label{subtr}
\end{align}
which reconstructs the amplitude in terms of its residues and its values at
additional points $a_i$ different from $z_i$.
This is a generalization of formula first written in this context in
\cite{Benincasa:2011kn}
and further discussed in \cite{Feng:2011jxa} where $a_i$ are chosen to be roots
of $A_n(z)$.

The other option is to use the all-line shift, i.e. deforming all
external momenta. This was inspired by the work by Risager
\cite{Risager:2005vk} and recently used for studying the on-shell
constructibility of generic renormalizable theories in
\cite{Cohen:2010mi}. This approach will be useful for our purpose.

\section{Semi-on-shell amplitudes}

The Lagrangian of the $SU(N)$ non-linear sigma model can be
written as
\begin{equation}
{\cal L} = \frac{F^2}{4}\text{Tr}\left( \partial_\mu U\,\partial^\mu
U^\dagger\right)\,,\label{Lagr}
\end{equation}
where $F$ is a constant and $U\in SU(N)$. In the most common
exponential parametrization $U={\rm exp}(i\phi/F)$ where
$\phi=\sqrt2\phi^a t^a$. The $t^a$s are generators of $SU(N)$ Lie
algebra normalized according to $\text{Tr}(t^a t^b) = \delta^{ab}$.
Note that for $N=2$, (\ref{Lagr}) is a leading ${\cal O}(p^2)$ term
in the Lagrangian for the Chiral Perturbation Theory
\cite{Gasser:1983yg}, which provides a systematic effective field
theory description for low energy QCD with two massless quarks. In
this case $\phi^a$ represent the pion triplet.

For calculations of on-shell scattering amplitudes within this model
we use stripped amplitudes $A_n(p_1,\dots p_n)$. The Lagrangian
(\ref{Lagr}) contains only terms with the even number of $\phi$,
therefore $A_{2n\psl1}=0$ and only $A_{2n}$ are non-vanishing. It is
easy to show that it makes no difference whether we use $SU(N)$ or
$U(N)$ symmetry group because the $U(1)$ piece decouples
\cite{future}. For our purpose it is convenient to use Cayley
parametrization of $U(N)$ non-linear sigma model,
\begin{equation}
U = \frac{1+\frac{i}{2F}\phi}{1-\frac{i}{2F}\phi} = 1 + 2
\sum_{n=1}^\infty \left(\frac{i}{2F}\phi\right)^n.
\end{equation}
Plugging for $U$ into (\ref{Lagr}) we get an infinite tower of terms
with two derivatives and an arbitrary number of $\phi$. This is
common for any parametrization, however, in this parametrization,
the stripped Feynman rule for the interaction vertex is particularly
simple,
\begin{equation}
V_{2n\psl1} = 0, \quad V_{2n\psl2} =
\left(\frac{-1}{2 F^2}\right)^n \left(\sum_{i=0}^np_{2i\psl1}
\right)^2.\label{vertex}
\end{equation}
It is easy to see that the shifted amplitudes $A_n(z)\approx z$
for $z\rightarrow\infty$. Without additional information on the values at two
points $a_i$ the relation (\ref{subtr}) cannot be used.
Therefore, we will follow different strategy to determine
$A_n(z)$ recursively.

Let us define a semi-on-shell current
\begin{equation}
J_n^{a_1,a_2,\dots a_n}(p_1,\dots p_n) = \langle 0|
\phi^a(0)|\pi^{a_1}(p_1)\dots \pi^{a_n}(p_n)\rangle
\end{equation}
as a matrix element of the field $\phi^a(0)$ between vacuum and the
$n$-particle state $|\pi^{a_1}(p_1)\dots \pi^{a_n}(p_n)\rangle$. The
momentum $p_{n\psl1}$ attached to $\phi^a(0)$ is off-shell
satisfying $p_{n\psl1} = - \sum_{j=1}^n p_j= -P_{1n}$. At the tree-level the
current can be written as a sum of stripped currents
$J_n(p_{\sigma(1)}\dots p_{\sigma(n)})$ as
\begin{align}
J_n^{a_1,a_2,\dots a_n}(p_1,\dots p_n)&=\\
&\hspace{-1.5cm}\sum_{\sigma/{\bf Z}_n} {\rm
Tr}(t^{a}t^{a_{\sigma(1)}}\dots
t^{a_{\sigma(n)}})J_n(p_{\sigma(1)}\dots p_{\sigma(n)})\,.\nonumber
\end{align}
The on-shell amplitude
$A_{n\psl1}(p_1,\dots p_{n\psl1})$ can be extracted from
$J_n(p_1,\dots p_n)$ by means of the LSZ formulas
\begin{equation}
A_{n\psl1}(p_1,\dots p_{n\psl1}) = - \lim_{p_{n+1}^2\rightarrow
0}\,p_{n\psl1}^2J_n(p_1,\dots p_n).\label{JtoA}
\end{equation}
The one particle states are normalized according to $J_1(p)=1$. Note
that $J_{2n}=0$ in agreement with $A(p_1\dots p_{2n\psl1})=0$ via
(\ref{JtoA}). For currents $J(1,\ldots,n)\equiv J_n(p_1,\dots p_n)$
we can write generalized Berends-Giele recursion relations
\cite{Berends:1987me} (n.b. $P_{ab}=\sum_{k=a}^b p_k$),
\begin{align}
J(1,\dots n) &= \frac{i}{p_{n\psl1}^2} \sum_{m=3}^n
\sum_{j_0<j_1<\ldots<j_m} \!\!\!\!\!\!i
V_{m\psl1}(P_{j_0j_1},\dots,-P_{1n})
\notag\label{Berends}\\ &
 \times \prod_{k=0}^{m-1}
J(j_k\psl1,\dots,j_{k\psl1})\,,
\end{align}
where $j_0=0$ and $j_m=n$. This equation can be equivalently
graphically represented as
\vspace{-0.28cm}
$$\includegraphics[scale=0.65]{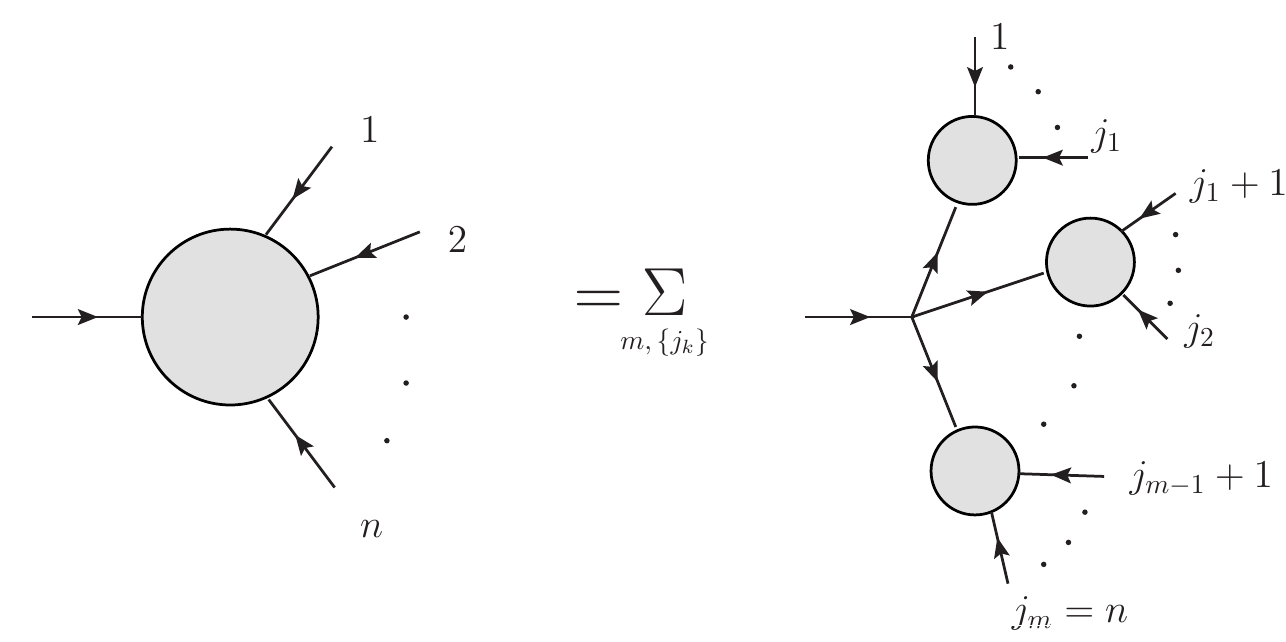}$$
The right hand side is a sum of products of lower point currents
with Feynman vertices (\ref{vertex}). The current $J_n$ is obviously
a homogeneous function of momenta of degree 0. It is not cyclic
because there is a special off-shell momentum $p_{n\psl1}$. Note,
however, $J_n$ is unphysical object and can be different in
different parametrizations. From now on we will use only Cayley
parametrization where it has interesting properties under the
re-scaling of all even or all odd on-shell momenta. Namely for
$t\rightarrow 0$:
\begin{align}
&J_{2n\psl1}(tp_1,p_2,tp_3,\dots p_{2n},tp_{2n\psl1}) =
O(t^2),\label{scaleOdd}\\
&J_{2n\psl1}(p_1,tp_2,p_3,\dots tp_{2n},p_{2n\psl1}) \rightarrow
\frac{1}{(2F^2)^n}\label{scaleEven}.
\end{align}
We postpone the detailed discussion to \cite{future}. The proof is
by induction using Berends-Giele recursion relations
\cite{Berends:1987me} which are more suitable for this purpose than
the analysis of Feynman diagrams used to show scaling properties of
Yang-Mills theory and gravity in \cite{ArkaniHamed:2008yf}.

\section{New recursion relations}

The scaling properties (\ref{scaleOdd}) and (\ref{scaleEven}) are
our guide for finding recursion relations for $J_{2n+1}$. Let us define
the complex deformation of the current $J_{2n+1}(z)$:
\begin{equation}
J_{2n+1}(z) \equiv J_{2n+1}(p_1, z p_2,\dots, zp_{2n}, p_{2n\psl1})\,,
\end{equation}
i.e. the momenta are shifted according to
\begin{equation}\label{pzdef}
p_{2k}(z)=zp_{2k}\,,\quad p_{2k\psl1}(z)=p_{2k\psl1}\,.
\end{equation}
Note that the momentum conservation is hold because the off-shell
momentum $p_{2n\psl2}=-\sum_{k=1}^{2n\psl1} p_k$ becomes also
shifted. In the limit $z\rightarrow0$ using (\ref{scaleEven}) we get
\begin{equation}
\lim_{z\rightarrow0}J_{2n+1}(z) = \frac{1}{(2F^2)^n}.\label{zeroZ}
\end{equation}
On the other hand for $z\rightarrow\infty$ we get as a consequence of
homogeneity and (\ref{scaleOdd}) the current $J_{2n+1}(z)$ vanishes like
\begin{equation}
J_{2n+1}(z) = O\left(\frac{1}{z^2}\right)\label{largeZ}
\end{equation}
and we can use the standard BCFW recursion relations to reconstruct
it from its poles.
The singularities of the physical current $J_{2n+1}(1)$ are determined by
condition
$P_{ij}^2=0$ which implies the following condition for the poles of
$J_{2n+1}(z)$
\begin{equation}
P_{ij}^2(z) = (zp_{ij}+q_{ij})^2=0\,,\label{polesZ}
\end{equation}
where $j-i$ is even and we have decomposed $P_{ij} = p_{ij}+q_{ij}$ where
$p_{ij}$ and
$q_{ij}$ is the sum of even and odd momenta respectively between $i$
and $j$,
\begin{equation}
p_{ij} = \sum_{i\leq 2k\leq j} p_k, \qquad\quad q_{ij} = \sum_{i\leq
2k\psl1\leq j} p_{2k\psl2}.
\end{equation}
For $j-i > 2$ we find two solutions of (\ref{polesZ}), namely
\begin{equation}
z_{ij}^\pm = \frac{-(p_{ij}\cdot q_{ij})\pm \sqrt{(p_{ij}\cdot
q_{ij})^2 - p_{ij}^2q_{ij}^2}}{p_{ij}^2}.\label{solZ}
\end{equation}
For the special case of three-particle pole, $j-i=2$, either
$q_{ij}^2=0$ or $p_{ij}^2=0$. For the first case $z_{ij}^+=0$ and
the corresponding residue does vanish, ${\rm Res}(J_{2n+1},z_{ij}^+)=0$,
while $z^-_{ij} =-2(p_{ij}\cdot q_{ij})/p_{ij}^2$.
In the second case there is only one solution of (\ref{polesZ})
$z_{ij} = - q_{ij}^2/2(p_{ij}\cdot q_{ij})$.

Let us denote a generic solution of (\ref{polesZ}) by $z_P$. Then
the internal momentum $P_{ij}(z_P)$ is on-shell, therefore the
current $J_{2n+1}(z)$ factorizes into the product of lower-point
semi-on-shell current $J_{m_1}$ and the on-shell amplitude
$M_{m_2}$. 
Residues at the poles $z_{ij}^\pm$ are given by
\begin{multline}
{\rm Res}(J_{2n+1},z_{ij}^\pm) =
\mp [p_{ij}^2(z_{ij}^+-z_{ij}^-)]^{-1} M_{ij}(z_{ij}^\pm) \\
\times J_{2n\mi j\psl i +1}(p_{1}(z_{ij}^\pm),\ldots,P_{ij}(z_{ij}^\pm),\ldots
,p_{2n+1}(z_{ij}^\pm))\label{Res}
\end{multline}
or graphically by
$$\includegraphics[scale=0.6]{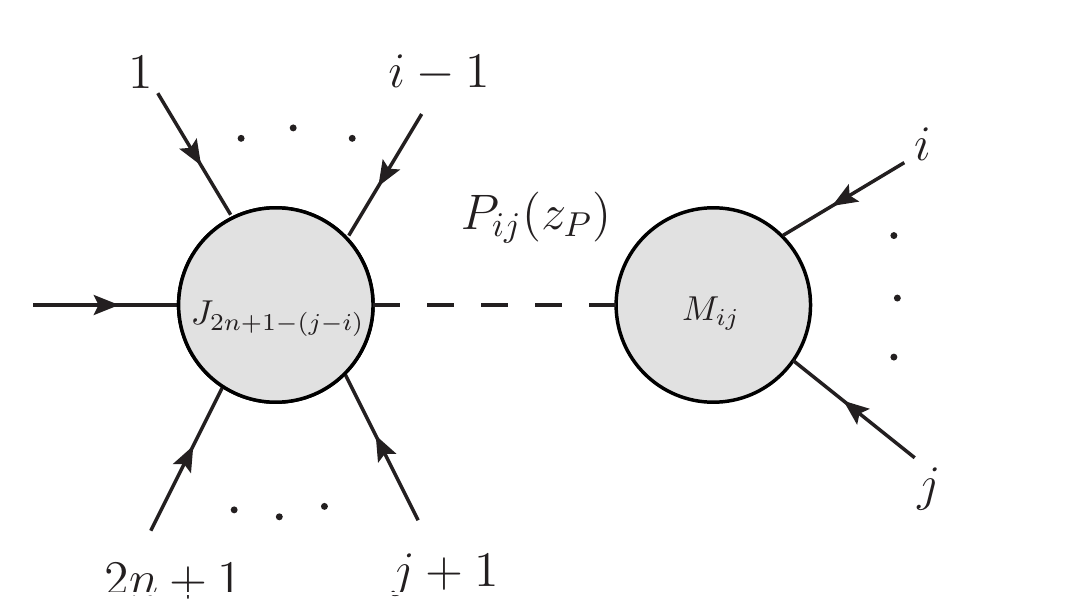}$$
In this formula  $M_{ij}(z) = P_{ij}^2(z)J_{j\mi i\psl1}(p_i(z),\dots
p_j(z))$. In the case of single solution $z_{ij}$ the residue is
given by the similar formula where $\mp
[p_{ij}^2(z_{ij}^+-z_{ij}^-)]^{-1}$ is replaced by $[2(p_{ij}\cdot
q_{ij})]^{-1}$. 
  
Because of (\ref{largeZ}) we can write
\begin{equation}
J_{2n+1}(z) = \sum_{z_P}\frac{{\rm
Res}(J_{2n+1},z_P)}{z-z_P}. \label{recursion}
\end{equation}
The residues ${\rm Res}(J_{2n+1},z_P)$ can be determined recursively
from (\ref{Res}) as in the case of BCFW recursion relations.
However, there is one difficulty. In the boundary case $i=1$,
$j=2n+1$ the equation (\ref{Res}) for residue ${\rm
Res}(J_{2n+1},z_{1,2n+1}^\pm)$ contains a current
$J_{2n+1}$ on the right hand side and therefore we
can not express it using lower point currents. The solution to this
problem is to use two extra relations. The first is the residue
theorem: because of the asymptotic behavior (\ref{largeZ}) the
residue at infinity vanishes and the sum of all residues is zero,
\begin{equation}
\sum_{z_P}{\rm Res}(J_{2n+1},z_P) = 0\,,\label{eq1}
\end{equation}
while the second one is the scaling property (\ref{zeroZ}) for
$z\rightarrow0$ together with (\ref{recursion})
\begin{equation}
\sum_{z_P}\frac{{\rm
Res}(J_{2n+1},z_P)}{z_P} = -\frac{1}{(2F^2)^n}.\label{eq2}
\end{equation}
Denoting $z_\pm = z_{1\,2n\psl1}^\pm$ and solving for ${\rm
Res}(J_{2n+1},z_\pm)$ from (\ref{eq1}) and
(\ref{eq2}) in terms of all other residues we can rewrite
(\ref{recursion}) in the form
\begin{align}
J_{2n+1}(z)& =
\frac{q_{1,2n+1}^2}{P_{1,2n+1}(z)^2}\frac{1}{(2F^2)^n} \notag\\
&\hspace{-0.65cm}
+{\sum_{z_P}}'\Bigl[\frac{z_+z_-}{(z-z_+)(z-z_-)}\frac{{\rm
Res}(J_{2n+1},z_P)}{z_P}\label{recursion2}\\ &\hspace{-0.65cm}-
z\frac{{\rm Res}(J_{2n+1},z_P)}{(z-z_+)(z-z_-)} +
\frac{{\rm
Res}(J_{2n+1},z_P)}{z-z_P}\Bigr]\,,\notag
\end{align}
where the sum is over all solutions of (\ref{polesZ}) with the
exception of $z_\pm$. The residues on the right-hand side depend
only on lower point currents via (\ref{Res}). The physical case is
$z=1$ and the on-shell amplitude $A_n(p_1,\dots p_n)$ can be
obtained from $J_n(1)$ using the limit (\ref{JtoA}). Interestingly,
even the fundamental 4pt case, i.e. the current $J_3$ is included  in
the equation (\ref{recursion2}) (here the sum is empty). Notice a very
important difference
between our recursion relations and the original Berends-Giele
formula (\ref{Berends}): we construct the amplitude recursively from
the 4pt formula via BCFW while (\ref{Berends}) uses critically the
Lagrangian and the infinite tower of terms in the expansion of
(\ref{Lagr}).

Detailed discussion of these results including the double soft-limit
formula and the proof of Adler's zeroes for stripped amplitudes
$A_n$ will be discussed in \cite{future}.

\medskip

\medskip

\subsection*{Conclusion and outlook}

We found the recursion relations for on-shell scattering amplitudes
of Goldstone bosons in the $SU(N)$ non-linear sigma model. We
defined a semi-on-shell current $J_n$ and used the Berends-Giele
recursion relations to prove its special scaling properties. This
allowed us to apply a particular all-line shift together with BCFW
construction to find the current recursively from the simplest
three-point case. The on-shell amplitude was then obtained from a
trivial limit when the off-shell momentum in $J_n$ became on-shell.

The existence of such recursion relations for effective theory gives
an evidence that on-shell methods can be used for much larger
classes of theories than has been considered so far. It also shows
that this theory is very special and deeper understanding of all its
properties is still missing. For future directions, it would be
interesting to see if the construction can be re-formulated purely
in terms of on-shell scattering amplitudes not using the
semi-on-shell current. Next possibility is to focus on loop amplitudes.
As was shown in \cite{ArkaniHamed:2010kv} the loop integrand can be
also in certain cases constructed using BCFW recursion relations, it
would be spectacular if the similar construction can be applied for
effective field theories.

\acknowledgments We thank Nima Arkani-Hamed and David McGady for
useful discussions and comments on the manuscript. JT is supported
by NSF grant PHY-0756966. This work is supported in part by projects
MSM0021620859 of Ministry of Education of the Czech Republic and
GAUK-514412.

\end{document}